\documentclass[acmlarge]{acmart}
\AtBeginDocument{%
  \providecommand\BibTeX{{%
    \normalfont B\kern-0.5em{\scshape i\kern-0.25em b}\kern-0.8em\TeX}}}

\setcopyright{acmlicensed}
\acmJournal{IMWUT}
\acmYear{2023} \acmVolume{7} \acmNumber{3} \acmArticle{108} \acmMonth{9} \acmPrice{15.00}\acmDOI{10.1145/3610903}

\usepackage{multicol}
\usepackage{multirow}
\usepackage{siunitx}
\usepackage{subcaption}
\usepackage{tabularx}

\def\markup{0}

\if\markup1
\newcommand{\revision}[1]{{\leavevmode\color{blue}#1}}
\usepackage{soul}
\usepackage[normalem]{ulem}
\else
\newcommand{\revision}[1]{#1}
\newcommand{\st}[1]{}
\newcommand{\sout}[1]{}
\fi

\begin{document}

\title[]{Exploring the Opportunities of AR for Enriching Storytelling with Family Photos between Grandparents and Grandchildren}


\author{Zisu Li}
\email{zlihe@connect.ust.hk}
\orcid{0000-0001-8825-0191}
\affiliation{
  \institution{The Hong Kong University of Science and Technology}
  \country{Hong Kong SAR, China}
}

\author{Li Feng}
\email{lfeng256@connect.hkust-gz.edu.cn}
\orcid{0000-0002-6198-0896}
\affiliation{
  \institution{The Hong Kong University of Science and Technology (Guangzhou)}
  \city{Guangzhou}
  \country{China}
}

\author{Chen Liang}
\email{liang-c19@mails.tsinghua.edu.cn}
\orcid{0000-0003-0579-2716}
\affiliation{
  \institution{Tsinghua University}
  \country{Beijing, China}
}

\author{Yuru Huang}
\email{yhuang760@connect.hkust-gz.edu.cn}
\orcid{0009-0003-1210-2194}
\affiliation{
  \institution{The Hong Kong University of Science and Technology (Guangzhou)}
  \city{Guangzhou}
  \country{China}
}

\author{Mingming Fan}
\authornote{Corresponding Author}
\email{mingmingfan@ust.hk}
\orcid{0000-0002-0356-4712}
\affiliation{
  \institution{The Hong Kong University of Science and Technology (Guangzhou)}
  \city{Guangzhou}
  \country{China}
  }
\affiliation{
  \institution{The Hong Kong University of Science and Technology}
  \city{Hong Kong SAR}
  \country{China}
}

\renewcommand{\shortauthors}{Li et al.}

\begin{abstract}

Storytelling with family photos, as an important mode of reminiscence-based activities, can be instrumental in promoting intergenerational communication between grandparents and grandchildren by strengthening generation bonds and shared family values. Motivated by challenges that existing technology approaches encountered for improving intergenerational storytelling (e.g., the need to hold the tablet, the potential view detachment from the physical world in Virtual Reality (VR)), we sought to find new ways of using Augmented Reality (AR) to support intergenerational storytelling, which offers new capabilities (e.g., 3D models, new interactivity) to enhance the expression for the storyteller. We conducted a two-part exploratory study, where pairs of grandparents and grandchildren 1) participated in an in-person storytelling activity with a semi-structured interview 2) and then a participatory design session with AR technology probes that we designed to inspire their exploration. Our findings revealed insights into the possible ways of intergenerational storytelling, the feasibility and usages of AR in facilitating it, and the key design implications for leveraging AR in intergenerational storytelling.
  
\end{abstract}





\maketitle

\section{Introduction}
\revision{Reminiscence activities have been proved to be an effective way to improve intergenerational communication and relationships between grandparents and grandchildren \cite{flottemesch2013,momentmeld,10.1145/2598784.2602769,10.1145/3415169}. Grandparents share their personal experiences with their grandchildren as a means to preserving cultural connections \cite{10.1145/3555158} and fostering closer relationships and intimacy with their grandchildren \cite{flottemesch2013learning}. In addition, reminiscence activities are effective in fostering increased empathy among grandchildren for their grandparents \cite{gigliotti2005intergenerational,chung2009intergenerational} and in facilitating their acquisition of knowledge \cite{nicholson2009social}.}



\sout{The global aging trend has brought challenges to society, especially in elderly care. Older adults not only need support to deal with aging symptoms (e.g., physical disability, memory loss) but also social challenges (e.g., isolation, loneliness, disconnection from society, and lacking communication with their children) \cite{fingerman2011gets,long2000personality,doi:10.1080/03601277.2014.912454}. To alleviate social challenges, communication and companionship from children and grandchildren are considered an efficient alleviation that many older adults seek. Unfortunately, there is a decline in intergenerational communication and support \cite{10.1093/geronb/gbq009}, probably due to a mutual effect of many factors, like migration \cite{liaqat2021participatory}, stereotypes \cite{williams2013intergenerational}, and overaccommodation \cite{williams2013intergenerational,giles2011intergenerational}.}

\revision{\textit{Intergenerational storytelling}, especially through the help of \textit{old physical photos} as a medium or a trigger, is the primary mode of engagement in reminiscence activities across generations \cite{momentmeld,flottemesch2013,thompson2009,liaqat2021participatory}. The craft of storytelling relies heavily on three factors including the content (e.g., themes, consequences, etc.) of a story, the storyteller's presentation skills (e.g., oral languages, tone, facial expressions, body languages, etc.) and the organization strategies for key points in a story \cite{YODERWISE200337}. However, these three components make it challenging to engage in successful storytelling for intergenerational pairs due to differences in life experiences, communication approaches, and issues caused by generational gaps \cite{stargatt2022digital,tokunaga2021dialogue,doi:10.1080/01924788.2018.1448669}.} 
\revision{To enhance intergenerational storytelling, researchers have explored 2D technologies \cite{kucirkova_ss_with_ipad,west2007memento} and virtual reality (VR) \cite{10.1145/3275495.3275513} to enrich and visualize the intergenerational story's content or improve interactions between two generations. However, many existing 2D technologies for augmenting photos require storytellers to hold the device, which occupies one or both hands. Furthermore, storytellers must switch back and forth between physical photos and the augmented content on a tablet user interface, which limits immersion and may cause distractions \cite{aitamurto2018impact}. These interruptions in the interface, story elements (such as old photos), and the listener's attention can lead to a temporary loss of focus on the story and a breakdown in its coherence \cite{bavelas2000listeners}. }

\revision{Compared to 2D technologies, 3D technologies, such as VR or AR, allow storytellers and listeners to interact with story elements in an immersive environment, which may provide richer and multi-modal information exchange during storytelling \cite{aitamurto2018impact, bavelas2000listeners}. Compared to VR that blocks users from the physical world, AR can enrich story elements by augmenting a physical photo with digital content beside it, which allows storytellers and listeners to access both the physical and digital worlds simultaneously \cite{9875213,yilmaz2017using,zhou2004magic,calvi2020we,seifert2021use,app9173556,healey2021mixed,putra2022utilizing,benckendorff2018role}. Moreover, different from VR, AR allows storytellers and listeners maintain natural eye contacts and use natural gestures while engaging in the storytelling experience.}

\sout{Researchers have investigated new forms of intergenerational communication that both the older and younger generations could take advantage of, such as storytelling \cite{flottemesch2013}, social games\cite{hyde2000recognising}, and mentoring systems on social media\cite{nagai2014t}. Among these forms, photos have been frequently used as prompts to trigger proactive interactions\cite{momentmeld}, share personal experiences\cite{10.1145/2598784.2602769}, and promoting reminiscence \cite{10.1145/3415169}. In particular, photo-based storytelling has been shown to facilitate intergenerational communication, maintain cultural connections, and share family value \cite{flottemesch2013,thompson2009,liaqat2021participatory}. In addition to the traditional form of storytelling with physical photos, assistive technologies with smart devices (e.g., tablets \cite{kucirkova_ss_with_ipad}, tangible interfaces \cite{west2007memento}) have been designed to improve the interaction between generations during the storytelling. However, most of these technologies required users to hold the devices and interact on 2D interfaces (e.g., tablet), which could provide limited immersion and may cause distraction \cite{aitamurto2018impact} among the switching operations on the interfaces, story elements (e.g., old photos), and the listeners, leading to temporarily oblivion about the details of the story and destroy the coherence of the story \cite{bavelas2000listeners}. }

\sout{Emerging 3D interactive augmented reality (AR) devices, such as Microsoft Hololens, equip an optical see-through lens and are suitable for applications involving human-human and human-environment interactions (e.g., collaborative activities and communication \cite{1115083,doi:10.1207/S15327590IJHC1603_2}, interacting with physical storytelling prompts such as photographs\cite{10.1145/3351232}). For example, AR allows users to interact with physical objects and augmented digital content while still maintaining eye contact with others. This ability may provide rich and multimodal information exchange and allow for eye contact between storytellers and listeners during the storytelling process. Indeed, researchers recently began to explore the potential of AR for enabling intergenerational communication \cite{seifert2021use,app9173556,healey2021mixed,putra2022utilizing,benckendorff2018role}.}


Although AR has the potential to enrich photo-based intergenerational storytelling, which has been shown to be beneficial for grandparents and grandchildren \cite{10.1145/3301019.3323902,slotsmomento,10.1145/3313831.3376486}, little is known about how grandparents and grandchildren might want to leverage AR to enhance their storytelling experience. 
In this work, we took a first step to explore this problem space. As it remains unclear how grandparents and grandchildren engage in photo-based storytelling, we took an open-ended participatory design approach, similar to the ones used in prior work \cite{10.1145/3555158,liaqat2021participatory}, to first investigate their photo-based storytelling practices and then explore the opportunities of AR to augment this process. Specifically, we sought to answer the following two research questions (RQs):

\textbf{RQ1:} \revision{How do grandparents and grandchildren perform photo-based intergenerational storytelling?}

\textbf{RQ2:} How might AR help photo-based intergenerational storytelling between grandparents and grandchildren? 

To answer RQs, we invited grandparent-grandchildren pairs ($N=10$) to participate in a two-part mixed-method study. Specifically, in part one, we first invited grandparent and grandchild pairs to engage in an open \revision{in-person} \sout{in-situ} photo-based storytelling activity \revision{and then interviewed them to understand their storytelling experiences and challenges. In part two, following prior design probe based approaches \cite{riche2010peercare,10.1145/3290605.3300823}, we asked them to try out a series of six AR probes that we designed to inspire them to think about the possible ways in which AR could enhance their storytelling process.} \sout{grandparent-grandchildren pairs participated in a photo-based \revision{in-person} storytelling activity followed by a semi-structured interview.Part 1 aimed to elicit the practices of intergenerational storytelling and also helped the pairs get ready for Part 2, which was a pair-wise participatory design workshop prompted by six AR probes potentially useful in storytelling. After getting familiar with AR, participants tried each of the AR technology probes designed to inspire them to think about the potential usage of AR in their storytelling process} \revision{After trying out these probes, grandparent and grandchild pairs participated in a workshop to brainstorm and design possible }AR interactions and functions that could assist their intergenerational storytelling. 

\revision{By analyzing both the observations and interviews from the part one of the study, we} uncovered three types of content (e.g., family portraits and activities), two tendencies in presenting content (e.g., verbal-centered storytelling with limited visual or audio support), and two characteristics in the organization of stories (e.g., linking stories with a relevant person) in their photo-based storytelling processes. Furthermore, \revision{by analyzing the design outputs and feedback from the part two of the study,} we uncovered seven ways in which AR could improve photo-based intergenerational storytelling. Lastly, we present the design considerations and future direction for improving intergenerational storytelling. 

In sum, we make three contributions:
1) We identified \revision{how grandparents and grandchildren perform photo-based intergenerational storytelling in terms of story content, presentation, and organization. }2) We uncovered seven ways how grandparents and grandchildren would want to leverage AR to assist their intergenerational storytelling. 3) We proposed design implications and possible ways to integrate the implications for designing AR-enabled photo-based intergenerational storytelling.

\section{Related Work}


\revision{We first show the importance of intergenerational storytelling as a reminiscence activity to foster better intergenerational connection and elaborate on the challenges in supporting storytelling between grandparents and grandchildren. Subsequently, we show the unique advantages of AR compared to other digital technologies (e.g., tablet-based approaches and VR) in facilitating intergenerational storytelling. Next, We show AR's potential in enhancing overall storytelling experiences and highlight the gap for leveraging AR to support intergenerational storytelling.}


\sout{
The growth of the aging population has become a global trend over the past decades \cite{bloompopulation,2016258,jyl2004}. Statistics showed a percentage ranging from 10 to 43\% of the social isolation rate among community-dwelling older adults with an ascending trend \cite{nicholson2010predictors,smith2009predicting}. Older adults who suoffered from social isolation often report a lack of communication, a lack of engagement in social activities, and attenuating social connections, especially with their family members \cite{healthcare6010024,taylor2018social}, leading to many psychological and physical health problems \cite{newall2019loneliness, nicholson2009social, nicholson2012review}. Previous work also validated the crucial role of family communication for the mental health of older adults against depression, loneliness, and even suicide \cite{matheson2020intergenerational,giles2008perceptions,kim2011intergenerational}.}
\sout{Intergenerational communication between individuals from different age groups has shown wide benefits for older adults from physical \cite{tam2014intergenerational, doi:10.1080/03601277.2014.912454,10.1145/3381002}, mental \cite{xu2016improving, doi:10.1080/108107397127833}, and social psychological aspects \cite{xu2016improving}. For instance, intergenerational communications with learning activities have been proven effective in maintaining cognitive functioning and capability \cite{ardelt2000intellectual, boulton2006learning,dench2000learning,10.1145/3369821}. In addition, for the younger generation, intergenerational communication could bring them increased social acceptance \cite{femia2008intergenerational}, greater willingness to help, and increased empathy for older adults \cite{gigliotti2005intergenerational}.}

\subsection{\revision{Photo-based Intergenerational Storytelling in Reminiscence-based Activities}}
Reminiscence-based activity is defined as an activity that involves recalling and revisiting a memory \cite{hallford2016brief}. It may involve the utilization of videos, music, pictures, and other objects that hold significant personal meaning for an individual. Among different types of reminiscence-based activities, reminiscence storytelling sits at the nexus of personal and collective memories, where the older generation shares stories about their living experiences, familial past, and values with the younger generation \cite{fivush2008intergenerational}. \revision{It has been proven to benefit both the younger and older generations \cite{hewson2015engaging,manchester2015digital, MERRILL201672}. For instance, the younger generation could gain knowledge and life experiences from the older generation, and the older generation could improve their life satisfaction and understanding of the younger generations \cite{hewson2015engaging}. It also provides an instructive platform to help family members build self-identity and be conducive to inheriting positive family legacies \cite{manchester2015digital}. In addition, a literature review on intergenerational narratives found that intergenerational narratives could influence individuals' psychosocial development \cite{MERRILL201672}. }
\sout{Previous work has explored the benefits of storytelling for fostering intergenerational relationships \cite{10.1145/2212776.2223698,doi:10.1080/03637750500322453,doi:10.1080/15267430903401441} and improving the well-being of both elder and younger generations \cite{MERRILL201672,adler2016incremental,hewson2015engaging,10.1145/1551788.1551875}. The younger generation gained knowledge and life experiences of the older generation, and the older generation gained an increase in their life satisfaction and a better understanding of the younger generations \cite{hewson2015engaging} and both generations enjoyed the communication and storytelling processes. Intergenerational storytelling also provides an instructive platform to help family members build self-identity and be conducive to inheriting positive family legacies\cite{manchester2015digital, MERRILL201672}. A literature review on intergenerational narratives \cite{MERRILL201672} found that intergenerational narratives could influence individuals' psychosocial development.}

\revision{Photo-based intergenerational storytelling is a powerful and pervasive method in reminiscence activities \cite{flottemesch2013}. A family photo is an essential vehicle for conducting multimedia digital storytelling as it could trigger instructive storytelling in a family \cite{gonzalez2012photo} and reminiscence \cite{flottemesch2013}. In addition, it could also provide opportunities to provoke interactions between the listener and the storyteller\cite{slotsmomento}, like sharing life experiences among family members. Considering the unique and powerful role of photo-based storytelling in reminiscence-based activities for both older and younger generations, our work focused on the photo-based storytelling process between grandparents and grandchildren with exploring ways to improve the experience in it.}


\subsection{\revision{Challenges in} Photo-based Intergenerational Storytelling}
Although Photo-based intergenerational storytelling was proven to be beneficial, \revision{
it can be challenging for grandparents to handle the three aspects of intergenerational storytelling (i.e., content, presentation skills and organization strategies \cite{YODERWISE200337}) to make it engaging for grandchildren due to differences in their life experiences, communication approaches, and other issues caused by generational gaps \cite{stargatt2022digital,tokunaga2021dialogue,doi:10.1080/01924788.2018.1448669}.}

\revision{The content of a well-developed story is expected to focus on key objects and events while leaving out minor details that might hinder story development \cite{10.3115/980491.980493}.} A study compared the contents and distribution of life story memories between the younger and older generation \cite{bohn2010generational}. It was found that the older generation often focuses on some historical events which the younger generation did not experience. This might result in difficulty in understanding and requires more description from the older generation.

\revision{A well-organized narrative structure could make the story easier to comprehend. For instance, Liem et al. \cite{liem2020structure} found that well-organized stories could significantly enhance the listener's understanding. The role of the organization in storytelling could also be generalized to storytelling in online posts. 
However, the organizational strategies of the older generation may confuse the younger generation. An analysis of narratives collected from two different generations revealed significant differences in the hierarchical and syntactic structure between the older and the younger generations \cite{kemper1990telling}.}

\revision{In addition, the presentation skills of the storytelling process could highly impact the user experience of the storyteller and the listener, such as the comprehension of the story and engagement. For instance, using different tones would affect the empathy arousal of the listener \cite{henningsen2019digital}. An analysis of the memorizing effect of narratives told in different presentation styles showed significant differences between the older and younger generations \cite{smith1983adult}. } However, the older generation was found to show a decrease in presentation skills, including the incorporation of visual aids when communicating with audiences \cite{worrall1998evaluation,10.1145/3555158}. 
\revision{Despite the importance of the three elements in storytelling, it remains largely unknown how grandparents and grandchildren engage in photo-based storytelling, including their strategies and challenges. By comprehending this gap, we can identify potential opportunities for leveraging technologies. Thus, we were motivated to investigate this problem as our first research question.}
\sout{Story-me \cite{10.1145/3301019.3323902} provided a cellphone application for the younger adults and a slots-machine-like device for the older adults to display and share family memento stories.} 
\sout{There are also tools focused on augmenting interactions based on photos between older and younger generations. For instance, Cueb \cite{golsteijn2013facilitating} presented a set of interactive digital photo cubes with which parents and teenagers can explore individual and shared experiences and are triggered to exchange stories. Momentmeld \cite{momentmeld} developed an AI-powered, cloud-backed mobile application that serves users within their daily routines of photo-taking and interaction by juxtaposing multiple semantically similar photos, each from different generations. In sum, prior work primarily focused on facilitating photo-based story sharing in a remote \cite{10.1145/3301019.3323902,slotsmomento,momentmeld} or asynchronous \cite{slotsmomento} scenario while few explored in-person story sharing experience, where modalities, effects, and designs for in-person interaction between generations is worth further investigation. In this work, we were inspired to explore how to leverage emerging technology, such as AR, to enrich in-person storytelling experiences for both older and younger adults.}


\subsection{Digital Technologies to Facilitate Photo-based Intergenerational Storytelling}


\sout{Photos, as an important form of reminiscence, are communication mediators to trigger instructive storytelling in a family \cite{gonzalez2012photo} where it could open opportunities for sharing life experiences among family members.} 

\revision{Various digital technologies have been proposed to enhance photo-based intergenerational storytelling experience in aspects of facilitating photo-based interaction(e.g., photo preservation and sharing \cite{slotsmomento,10.1145/3313831.3376486}), enhancing asynchronous interactivities \cite{golsteijn2013facilitating,momentmeld}, and facilitating personal expressions \cite{sehrawat2017digital,hewson2015engaging}.
Among digital technologies used to enhance intergenerational storytelling, both tablet-based approaches and 3D technologies were invented to enhance intergenerational storytelling. FamilyStories \cite{10.1145/3313831.3376486} allows a person to record an audio story and send it to another family member. Story-me \cite{10.1145/3301019.3323902} provided a cellphone application for children and a slots-machine-like device for older adults to display and share family memento stories. Momentmeld \cite{momentmeld} developed an AI-powered, cloud-backed mobile application that serves users within their daily routines of photo-taking and interaction by juxtaposing multiple semantically similar photos, each from different generations. Nonetheless, the utilization of tablet-based methods presents a challenge as users are required to hold the device in their hands, potentially impeding them from simultaneously holding and manipulating physical artifacts, such as paper photos, to enhance their storytelling experience.  Moreover, users may need to split their attention between the tablet screen and the artifact to be held \cite{liu2012split}, which might disrupt the flow and consistency of the storytelling process. Thus, we chose more immersive technologies that could make users free their hands to make more interactions during storytelling.}

\revision{There is also prior research focused on using immersive technologies, such as VR and AR, to improve intergenerational storytelling. Thomas et al. provided an immersive and interactive VR experience that encourages participants to question their use of plastic \cite{10.1145/3275495.3275513}. Cutler et al. present their experiments and learning in authoring interactive VR narratives across different scenarios\cite{10.1145/3388767.3407319}. Compared to VR, which blocks users from the physical world, AR can enrich story elements by augmenting a physical photo with digital content beside it, which allows storytellers and listeners to access both the physical and digital worlds simultaneously \cite{9875213,yilmaz2017using,zhou2004magic,calvi2020we,seifert2021use,app9173556,healey2021mixed,putra2022utilizing,benckendorff2018role}. Thus, AR technology has the potential to foster collaboration, shared experience, and co-creation in storytelling, which motivated our choice of AR to facilitate the photo-based intergenerational storytelling process. }

\revision{Furthermore, we found most prior work primarily focused on facilitating photo-based story sharing in a remote \cite{10.1145/3301019.3323902,slotsmomento,momentmeld} or asynchronous \cite{10.1145/3313831.3376486,slotsmomento} scenario. However, facilitating the experience of in-person story-sharing activities is of equal or more importance in promoting intergenerational relationships and communication\cite{sehrawat2017digital}. Although some work investigated the usage of digital tools and applications (e.g., a digital story-making website) on commodity computers or tablets \cite{sehrawat2017digital,hewson2015engaging}, few of them explicitly identify the effects of modalities, effects, and designs of digital tools for in-person interaction for intergenerational storytelling. 
Considering the unique potential of AR for improving intergenerational storytelling, we sought to explore how grandparents and grandchildren leverage emerging AR technology to enrich in-person photo-based intergenerational storytelling experiences.}

\subsection{Potentials of AR for Enhancing Storytelling Performance and Experience}


Taking advantage of development in display techniques \cite{jang2019progress} and computer vision (CV) techniques (including SLAM \cite{mur2017orb}, hand tracking \cite{han2020megatrack}, and object detection \cite{wang2022yolov7}), modern head-mounted AR devices allow users to see and interact with virtual objects in the immersive environment, which served as an important medium to enrich the visual effects and interaction channels for storytelling \cite{yilmaz2017using}. Previous work demonstrated AR's potential in enriching the three key factors (content \cite{santano2018augmented,yilmaz2017using,brusaporci2017augmented}, organization \cite{braun2003storytelling,li2022interactive,yilmaz2017using}, and presentation \cite{yilmaz2017using,zhou2004magic}) for storytelling.

\revision{Regarding narrative content, AR technology allows storytellers to augment the story by overlaying virtual objects, characters, or information onto the physical world, providing a richer and more engaging narrative experience \cite{santano2018augmented,yilmaz2017using}. For example, Yilmaz et al. \cite{yilmaz2017using} presented to overlay virtual elements onto the real world to create diverse storytelling scenes for examining elementary students’ narrative skills and creativity. Further, AR brings the capability of rendering immersive and realistic environments, which can be used for the recurrence of historical objects and sites. For example, Brusaporci et al. \cite{brusaporci2017augmented} leverage AR technology to reconstruct tangible and intangible images of a historical site for historical storytelling. }

\revision{AR offers new ways for the storyteller to organize and develop the story \cite{braun2003storytelling}. With AR, storytellers can create branching narratives or interactive storytelling experiences where virtual avatars and dynamic elements in AR can affect the direction of the story. For example, Braun et al. \cite{braun2003storytelling} introduced a new interactive storytelling practice in collaborative augmented reality combining an audience participatory theatre and a morphological approach to storytelling. Li et al. \cite{li2022interactive} proposed a method to synthesize virtual characters and activities based on scene semantics to guide an interactive storytelling experience in augmented reality. By presenting different AR elements or scenarios based on user interactions, AR-enhanced storytelling could become more personalized and flexible in terms of narrative structure.}

\revision{AR techniques also have the potential to enhance the presentation skills and forms, typically by offering hints \cite{yilmaz2017using,li2022interactive}, new presentation channels \cite{santano2018augmented,zhou2004magic}, and multi-sensory feedback \cite{zhou2004magic}. For example, Zhou et al. \cite{zhou2004magic} combined AR technology with a tangible cube interface to provide multi-sensory experiences, including speech, 3D audio, 3D graphics, and touch, in storytelling. }

\revision{Although AR has the potential to enrich photo-based intergenerational storytelling, which has been shown to be beneficial for grandparents and grandchildren \cite{10.1145/3301019.3323902,slotsmomento,10.1145/3313831.3376486}, little is known about how grandparents and grandchildren might want to leverage AR to enhance their storytelling experience. In this work, we adopted a probe-based participatory design to investigate the potentials of AR in facilitating intergenerational storytelling experiences. We designed six AR interaction probes aimed to explore how to leverage AR to enhance intergenerational storytelling, mainly about the three key factors for storytelling.}

\section{Method}
\begin{figure}[tbh!]
    \centering
    \includegraphics[width=1\textwidth]{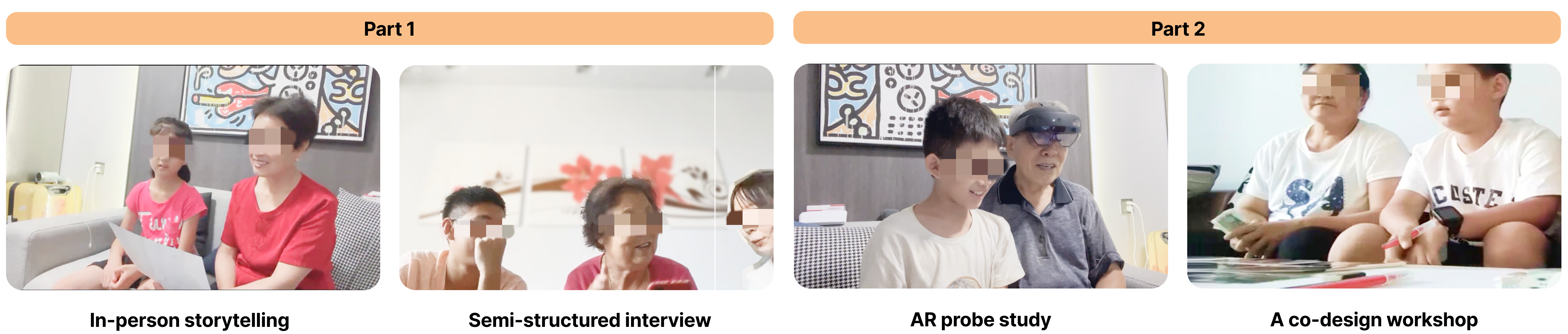}
    \caption{Procedure in the two parts of our user study: 1) In-person storytelling and a follow-up semi-structured interview; 2) pair-wise participatory design workshop with AR technology probes.}
    \Description{This figure contains the important sessions of our study which contains in-person storytelling, semi-structured interview, AR probe study, and a co-design workshop.}
    \label{fig:procedure}
\end{figure}

To answer our RQs, we conducted a two-part study as shown in Figure \ref{fig:procedure}: 1) an in-person storytelling session between grandparents and grandchildren, followed by a semi-structured interview and 2) a participatory design workshop including AR probe try-on activities and a co-design workshop. For simplicity, grandparent and grandchild are abbreviated as GP and GC respectively. 

\subsection{Participants}
\label{sec:participants}
We recruited 10 pairs of participants (20 in total) by word-of-mouth. The GP participants recruited had various occupations before they retired (e.g., teacher, sales, etc.). Each pair was in a grandparent-grandchild relationship. The age range of GPs was 61 to 78 ($mean= 70.5, SD=5.0$) and the age range of GCs was 12 to 22 ($mean=17.1, SD=3.0$). All GPs had no experience with AR or VR. Only one GC had experience with VR technologies. Most GP-GC pairs had frequent daily communication - 7 pairs communicated every day, and \revision{additional 2} pairs communicated at least once a week. Detailed demographic information of the participants is shown in Table \ref{tab:demographics}. For each pair of participants, the whole study took around 2 hours, and they together received 50 US dollars for compensation. This study was approved by the Research Ethics Board of our institution.

 \begin{table}
  \vspace{-0.3cm}
  \centering
  \caption{Demographics of the participant pairs.}~\label{tab:demographics}
    \vspace{-0.3cm}
    \begin{tabular}{|l|cc|cc|}
    \toprule
    \multirow{2}{*}{Pair ID} & \multicolumn{2}{c|}{Grandparent} &    \multicolumn{2}{c|}{Grandchild} \\
    \addlinespace[0.2em]\cline{2-5}\addlinespace[0.2em]
    & Gender & Age  & Gender & Age \\
    \midrule
    1 & F & 64 & F & 12  \\
    2 & F & 73 & M & 17 \\
    3 & M & 74 & F & 22 \\
    4 & M & 61 & M & 17 \\
    5 & M & 75 & M & 15\\
    6 & F & 78 & F & 22 \\
    7 & M & 66 & M & 14 \\
    8 & F & 69 & M & 17 \\
    9 & M & 72 & F & 19\\
    10 & M & 73 & M & 16 \\
    
    \bottomrule
  \end{tabular}
  \vspace{0.2cm}
  \Description{This table shows the demographics of the 10 participants, including ID, gender, and age.}
\end{table}

\subsection{Part 1: In-person Storytelling and Semi-structured Interview }

In Part 1, we aimed to answer RQ1. As shown in Fig \ref{fig:procedure}, we first conducted an in-person storytelling session with each GP-GC pair where two researchers observed their communication process during storytelling. Then we conducted a semi-structured interview with the participants to further understand the issues they faced in the three aspects (i.e., content, organization, and presentation) during their storytelling in the experiment and their previous intergenerational storytelling experiences.


\subsubsection{Preparation:} Before the study, we asked the GP to select five photos they would like to share the most and prepare stories of these photos in advance, aiming for high quality (e.g., the GP could recall more details and have a better organization) storytelling process. Notably, 
We requested GPs not to discuss the photos or their prepared stories with their GCs in advance or performed any rehearsal before our study to ensure that we could observe their most natural and intuitive communication processes during the study. 

\subsubsection{In-person storytelling:} The formal study was set up in the researcher's home, where participants sat on the sofa with a tea table and a television in front of them, to ensure that participants felt comfortable and relaxed to share their stories. After signing the consent form, participants were first given a brief introduction to the study. Then the GP was instructed to start telling the stories they prepared about the photos they brought in their preferred order. During the storytelling process, the GC was encouraged to have any interaction (e.g., asking questions or making comments) with the GP to promote communication and their understanding of the stories. Meanwhile, one researcher (R1) took notes of the contents of the stories that they (the researchers) were confused or surprised by. The whole storytelling process was recorded in video by the second researcher (R2). 

\subsubsection{Interview:} After the storytelling, R1 first asked the participants to describe their intergenerational relationship (e.g., communication frequency, intimacy, and storytelling experience), meta information of the photos (e.g., type, year, characters) as well as the storytelling characteristics (e.g., the organization of the story, the main topics, the narrative order, the ideas of GPs used to describe items) from the participants' perspectives, to have an overview of their storytelling activity. 

\subsubsection{Playback and Revision:} To further understand the challenges encountered by participants in storytelling, the researcher (R1) instructed both the GP and the GC to participate in a review phase, where the pair watched the video recording of their storytelling and R1's notes to reflect on this process. \revision{The researcher (R1) conducted interviews to inquire about participants' experiences, specifically whether any positive aspects or problems they encountered.
} Then the researcher (R1) asked some questions (e.g., Which part of the story do you think could be elaborated more? What is the order you used to arrange your story?) to encourage the GP and the GC together to think-aloud 1) the problems and difficulties they experienced during storytelling and 2) the possible improvements and modifications to the storytelling process (e.g., for better clarity or expressing more interesting points about the story). \revision{It is also important to note that we did not impose a requirement on participants to talk about the improvements for their storytelling process.} Meanwhile, R2 video-recorded the whole revision process and took notes of the key points from the think-aloud.

\subsection{Part 2: Participatory Design Workshop}
After part 1 of our study, the participants could recall more details in their daily storytelling process to get prepared for the following participatory design workshop which aimed at answering RQ2. In part 2 of our study, participants were instructed to participate in 1) an AR probe try-on activity
 and 2) a co-design workshop on AR interaction for intergenerational storytelling. We chose a probe-based study design instead of implementing and evaluating a full-functional prototype in this part because we wanted participants to contribute to the iterative improvement of an envisioned technology and work directly in the initial design circle with qualitative feedback \cite{10.1145/3491102.3517692,kaufman2018design}. Our AR probes were designed based on features mentioned in previous literature \cite{10.1145/2212776.2223698,doi:10.1080/03637750500322453,doi:10.1080/15267430903401441} that help augment intergenerational communication and existing AR interaction concepts.

\begin{figure}
    \centering
    \includegraphics[width=0.8\textwidth]{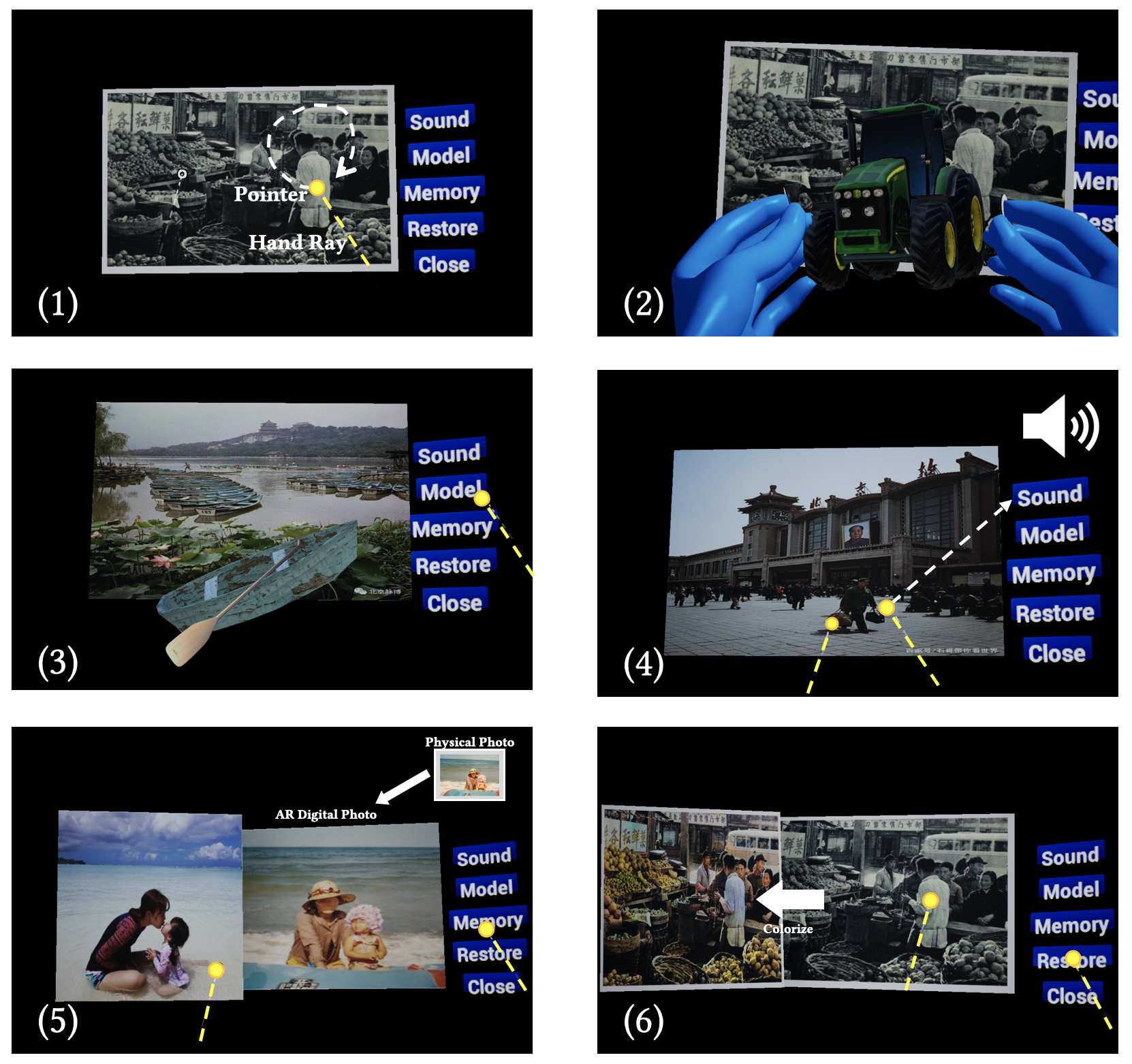}
    \caption{Illustrations of the AR probes used in our study. Hand rays and distant pointers in AR were highlighted in yellow. All the white arrows and icons were annotations and not displayed in AR. (1) Pointing - The user points and circles the man to highlight him during storytelling. (2) Object manipulation - The user rotates and resizes the tractor model by pinching and dragging with two hands. (3) Display of 3D models - When clicking the "Model" button, the 3D model of a key element in the photo is displayed. (4) Sound effects - When clicking the "Sound" button, photo-related sound effects or music will be played. (5) Reminiscence - Assuming physical family photos were scanned and converted into AR digital photos. When clicking the "Memory" button, a photo similar to (e.g., with a similar scene) the currently displayed photo shows up aside. (6) Photo restoration - When clicking the "Restore" button, a restored photo shows up beside the original old photo.}
    \Description{This figure contains six subsections, each demonstrating an AR probe: 1. pointing to the photo, 2. object manipulation, 3. display of 3D models, 4. background sounds, 5. Reminiscence, 6. Photo restoration.}
    \label{fig:probes}
\end{figure}

\subsubsection{AR Probe Design}

We devised six AR interaction probes and developed usage scenarios corresponding to each probe to demonstrate a preliminary design on how AR interactions could potentially be integrated into the storytelling process. 
We designed these probes based on existing AR interaction techniques (e.g., Hololens 2's interactions) and related literature. The goal of showing these probes to the participants was to give them first-hand experience of some possible usages of AR for storytelling and to inspire them to reflect on these probes and generate more ideas that they would feel AR could enhance their storytelling processes. 

Figure \ref{fig:probes} shows the six probes. Among them, two are general user interface (UI) operations in AR, and four are related to the storytelling process (ST). Next, we explain them in detail. \revision{These AR probes showcased the potential of AR in improving the storytelling process by leveraging two principal advantages. On one hand, in contrast to tablet based approaches, AR does not require users to hold the tablet with their hands. Thus, they can use their hands to hold the artifact (e.g., a paper photo) and manipulate it to enhance their storytelling experience. Moreover, they could also make use of hand gestures as people naturally do while conversing. Furthermore, while users may need to split their attention between the tablet and the artifact to be told, they can see and manipulate relevant digital content (e.g., 3D models) augmented around the artifact in one view through AR without splitting their attention. On the other hand, although VR can also augment digital content, it separates users from the physical world. Consequently, while telling physical photo-centered stories, unlike AR, VR does not enable users to easily see both the physical old photo and the augmented digital content simultaneously. Moreover, unlike AR, VR does not allow users to see their facial expressions and maintain eye contact, which is important for delivering an enjoyable storytelling experience. }

\textbf{(1) Pointing (UI):} Pointing is the basic UI interaction method in Hololens 2\footnote{https://docs.microsoft.com/en-us/windows/mixed-reality/design/point-and-commit}. A virtual ray is cast from the center of the palm (for both left and right hands), which intersects with a target object or widget and renders a dot-like cursor on its surface. The \revision{storyteller} could 1) point at an augmented object to attract the listener's attention during storytelling (similar to a laser pointer) and 2) \revision{select it by performing a pinch gesture} \sout{use pointing and pinching to perform indirect UI operations}. \revision{Moreover, one advantage of AR is that the storyteller can also use this pointing operation to point at an element in the physical photo.}

\textbf{(2) Object manipulation (UI):} The user could manipulate an object (e.g., a 2D UI widget or a 3D model) in the virtual environment with two types of interaction methods: indirect and direct. When manipulating distant objects, the user uses an indirect method with rays, where he could pinch and drag with both hands. For nearby objects, the user could directly manipulate the object by pinching, grabbing, and dragging with his hands. Panning, rotating, and scaling on an object are supported for both methods, which are consistent with Hololens 2's two manipulation models (``point and commit with hands'' \footnote{https://docs.microsoft.com/en-us/windows/mixed-reality/design/point-and-commit} and ``direct manipulation'' \footnote{https://docs.microsoft.com/en-us/windows/mixed-reality/design/direct-manipulation}). \revision{This AR probe offers a congruent means to manipulate physical and digital objects without needing to switch attention and learn a mapping between a 2D interface and the physical world as they would when using a 2D interface on a tablet.}

\textbf{(3) Display of 3D models (ST):} The 3D models of the elements related to the demonstrative photos (such as old artifacts and characters) were generated and put into the AR application preliminarily by the researchers. 
\sout{The user could simply press a "Model" button to evoke a 3D model associated with the current photo during the storytelling process. Informed by the prior work \cite{garzotto2006fate2} on a digital screen, we regarded the 3D display of the story-related models as a fundamental function for AR-assisted intergenerational storytelling.} 
 \revision{This probe was designed to simulate a scenario in which a storyteller wants to provide more details about a particular element in their story. 
For example, to help listeners better understand an artifact (e.g., a tractor used in the 1980s) that they have never seen, AR could render a 3D model of the tractor, which could be downloaded from a cloud assets database or be generated by generative AI algorithm and shown it beside the physical photo. Moreover, AI techniques could be designed to understand the conversation between the storyteller and the listener and trigger the appearance of such 3D models at the right time when the listener has difficulty understanding the artifact. We informed participants of this possibility in the future but asked them to press a ``Model'' button to trigger the appearance of the 3D model during the study.}
Furthermore, participants could also manipulate the model (e.g., dragging the model around or zooming in to view certain details) for better observation. 


\textbf{(4) Sound effects (ST):} Previous literature \cite{su2011photosense, chen2006tiling} found that playing background sounds related to photos could enhance the atmosphere of the story and make the listener and speaker feel immersed in the scene, thereby enhancing the effect of communication. Different types of background sounds for different photos were provided in the probe (e.g., old songs of the same age as the photo of a park, the sound of waves and seagulls for the photo on the beach, and the cries of hawkers for a market photo). \revision{This probe leverages AR's ability to provide spatial audio to augment the visual content of an old photo. In this way, we could seek participants' opinions of multi-sensory experiences for storytelling} 


\textbf{(5) Reminiscence (ST):} Inspired by MomentMeld \cite{momentmeld}, where the retrieval and display of the photos which has semantic similarities to the target photo in different domains (e.g., place-related, person-related, and family subjective) were proven beneficial for reminiscence, we designed a function to display photos with certain semantic relationships to the target photo in the AR environment. Specifically, we demonstrated the function with an example scenario as follows. We assumed that physical family photos were scanned and converted into AR digital photos. The currently displayed photo shows a scene where the grandma and the mother were playing on the beach. When the grandma pressed a button to trigger the reminiscence function, a photo of the mother and the daughter in a similar scene was displayed beside the original photo (on the left side). Such semantically related similar photos might enhance the association and reminiscence among the three generations. \revision{Taking inspiration from Axtell and Munteanu's work\cite{10.1145/3351232} advocating that digital tools should respect the role of paper pictures and integrate physical and digital picture interactions, this AR probe is intended to create a contrast between the physical photo and a digital augmented one, that is identified by current AI technologies, and to use such a contrast to make the storytelling more personally relatable and to trigger more interactions between the storyteller and the listener.}

\textbf{(6) Photo restoration (ST):} The quality (e.g., color and resolution) of the target photo was essential for the storytelling process. Many old photos suffer from quality problems, such as the degradation of color and stains. Restoration of old photos can be achieved with the assistance of advanced AI techniques, such as image colorization \cite{zhang2016colorful}, image super-resolution \cite{dong2015image}, and image completion \cite{drori2003fragment}. In our probes, we demonstrated the photo restoration function with an example of colorizing a black-and-white photo with a SOTA image colorization model 
\footnote{EnhanceFace Service from Alibaba. https://help.aliyun.com/document\_detail/159211.html}.\revision{Additionally, presenting old and restored photos in AR can also enhance the sensory experience of viewing the photos. Participants can interact with the photos in a more immersive and engaging way, such as zooming in to examine the details of the restored areas or moving around the photos to get a different perspective. This can also help to evoke emotions and memories associated with the photos, leading to more personal and reflective conversations.}

\begin{figure}[tbh!]
    \centering
    \includegraphics[width=\textwidth]{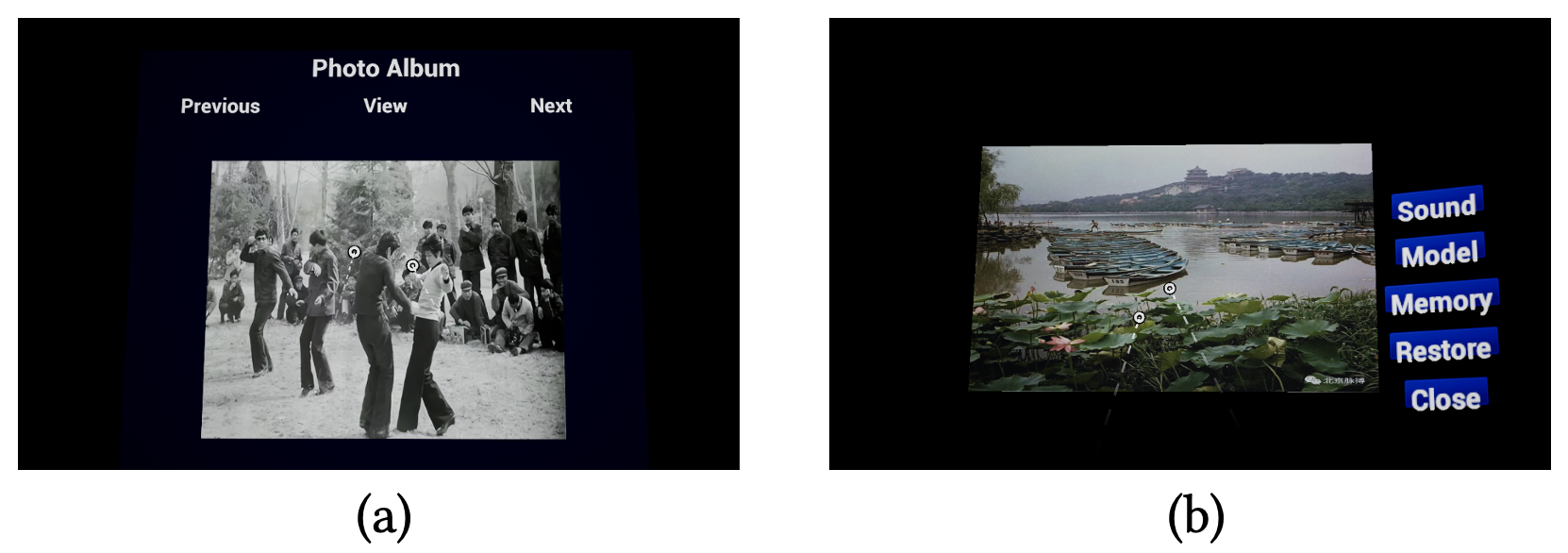}
    \caption{The UIs of our AR photo album app that integrates the design probes. (a) The photo album UI. (b) The photo widget.}
    \Description{This figure shows the UI of our AR photo album app. On the left is the photo album UI, which contains three buttons: previous, view and next. On the right is the photo widget, containing five buttons: sound, model, memory, restore and close.}
    \label{fig:ui}
\end{figure}

\textbf{Implementation:} We implemented an AR Photo Album app to integrate all the above-mentioned probes on a Hololens 2 using Unreal Engine 4 and Microsoft Mixed Reality Toolkit (MRTK). Fig. \ref{fig:ui} (a) shows the photo album UI where the user could browse different photos by clicking the "Previous" and "Next" buttons and spawn an individual photo widget (corresponding to the currently displayed photo, see Fig. \ref{fig:ui} (b)) by clicking the "Enter" button. The photo widget (Fig. \ref{fig:ui} (b)) consists of a photo display region at the center and a menu with five buttons - model, sound, reminiscence, restoration, and close - on the right, which correspond to the four functions for storytelling (4-6) above. The photo widget is movable and the user can spawn multiple widgets and lay them out in the virtual space simultaneously. \revision{It should be noted that all the effects of probes (e.g., 3D models, the sound effects, the semantically-related photo, the restoration for the old photo) were pre-designed and pre-imported to the app since we could control and optimize their quality and seamless integration within the storytelling process using AR. We will discuss the possible approaches for content generation in the following Discussion section. }

\subsubsection{AR Probe Try-ons}

\label{sec:probe}
Considering most participants were novices to AR technologies (see Section \ref{sec:participants}), we introduced an AR tutorial before the technology probe study to get them familiar with basic AR concepts, environments, and operations (e.g., pressing a button or operating a 3D object) in AR. Participants were instructed to wear an Hololens 2 and experience a built-in application - 3D Viewer\footnote{https://docs.microsoft.com/en-us/windows/mixed-reality/design/3d-modeling} - where the user could select 3D models or animations from a gallery and place them in the virtual space. They could also manipulate the 3D models with different hand gestures. The experimenter monitored the process and provided real-time instructions so that participants could successfully make each operation (open the virtual interface of 3D Viewer; choose a model; manipulate a model to put it in a suitable location to view; watch the animation of the model).

After participants got familiar with the AR environment and the basic interaction methods, they were instructed to experience the set of six AR probes for demonstrating concepts and interaction examples in intergenerational storytelling. 
The main purpose of experiencing these AR probes was to allow participants to think about how AR might be used to enhance storytelling, which would enable them to comment on these probes and inspire them to imagine new usages of AR in their storytelling processes. 
\revision{We considered two different settings of AR devices for presenting probes: using two AR devices, one for each participant, or using a single AR device. As a first step to exploring the possibility of using AR to augment intergenerational storytelling, we chose to use one AR device for two reasons. First, through our pilot studies with older adults and their grandchildren, we found that AR was a relatively new concept to them, and adopting AR into their storytelling process required some learning efforts. For example, it was challenging for them to set up a single AR headset and learn its operations. To reduce the complexity of the setup and focus on exploring the possibility of AR in helping their storytelling process, we chose to use just one AR headset for this initial research and let the storyteller try the AR device. Second, even if they could learn to set up two AR devices and operate two AR devices to tell a story with a physical photo,  there are more collaboration challenges, including potential operation conflicts or view management issues. }

\begin{figure}[tbh!]
    \centering
    \includegraphics[width=0.8\textwidth]{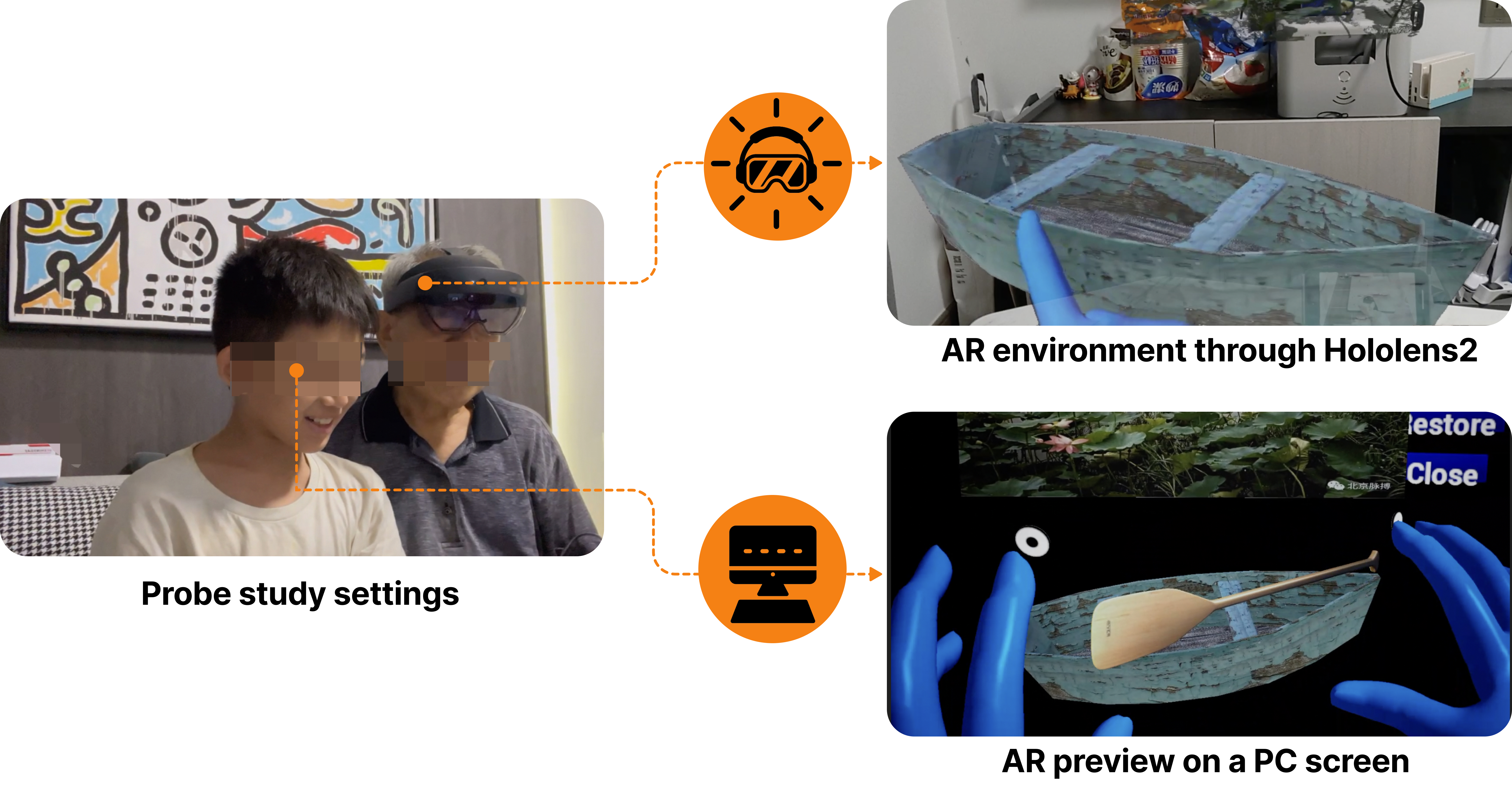}
    \caption{Probe study settings: GP wearing Hololens 2 and GC watching the display on PC screen from AR preview.}
    \Description{This figure demonstrates our study settings for a grandparent wearing Hololens 2 and the grandchild watching the PC screen from AR preview.}
    \label{fig:settings}
\end{figure}

During the probe activity procedure, the researcher first played a demo video to introduce the functions and usages of the probes with demonstrative scenarios and photos. Then the GP was invited to wear the AR glasses and try out the AR probes with the assistance of the researcher. Meanwhile, the GC monitored the GP's view on a PC screen and provided real-time feedback and comments to the GP. (See Fig. \ref{fig:settings}) After the GP got familiar with the interaction of the probes, the researcher encouraged him to simulate sharing the story or communicate with the GC using the probe techniques. After the GP's trying, the GC could wear the AR glasses and experience the AR probes if they found certain points in the probes of interest or for better clarification. When GCs tried the AR glasses, GPs monitored the view in the AR glasses through PC review and occasionally provided GCs with assistance and guidance.

\subsubsection{Co-designing AR Interactions for Intergenerational Storytelling}
\label{participatory_design}

 After the participants got familiar with the fundamental functions and operations of the AR environment and got to know some technology probes for AR storytelling, we instructed participants to take part in a participatory design workshop, where we 1) investigated how the technology probes could be used in AR storytelling, and 2) gained diverse and creative ideas and alternative solutions for improving AR storytelling applications. In the participatory design activity, participants were asked to think-aloud their ideas and conduct the end-user sketching \cite{end-user-sketch} to visualize their ideas. 

Participants were instructed to reflect on their experience of previous phases of the study and then think-aloud or visualize their ideas. The researcher first instructed the participants to depict the interactions and functions they would like to include in photo-based AR storytelling and draw them on the papers. If the GP cannot draw their ideas independently, they can ask the GC for assistance or simply write the ideas down, as prior work found that older adults are less willing to draw than the younger generation in participatory design \cite{xie2012connecting}. The researcher prompted this process by providing A3-sized white paper, pencils, crayons, stickers, and Lego bricks. Printed materials of technology probes in the previous phase were also provided to help participants refer. To encourage the participants to think more broadly, researchers asked prompting/provoking questions, which started with some modifications and further development of our technology probes. Afterward, the researcher interviewed participants about the reasons, the rationales, and the experiences about the design probes and the new ideas or new ways of using AR they came up with to improve the storytelling process.

\section{Data Analysis}


We collected approximately 20 hours of video recording from 10 pairs of participants in total, accompanied by auxiliary data including 1) field quick notes from the researchers, 2) photographs of the participatory design activities, 3) close-ups of the participatory design activities, 4) digital copies of the old photos brought by participants, and 5) digital copies of the sketches created from participatory design activities. The video recordings were transcribed into paragraphs using a commercial ASR system (iFLYTEK\footnote{https://www.iflyrec.com/zhuanwenzi.html}) and checked by the research team for correctness.


We performed semantic analysis \cite{braun2006using} to analyze our data. All of our co-authors first read the transcripts and supplementary materials (field notes, photographs, and sketches) to familiarize themselves with the data. Then, three co-authors open-coded the transcripts individually. The whole research team held weekly meetings to discuss the coding result, resolve disagreements, and update the code book. Based on the final coding results, the whole research team worked \revision{together} to iteratively group the codes and \revision{generate} the themes and subthemes, which are used to organize our findings.

\section{Findings}

\sout{Participants overall held positive attitudes toward the potential of AR technologies for facilitating their intergenerational storytelling process, especially about overcoming the difficulties they currently face, and contributed designs and insights into how AR could be leveraged to improve their storytelling process.}
\revision{We present our key findings of the two parts of our study to answer two RQs: 1) RQ1: Based on part 1 of our study (i.e., in-person storytelling and simi-structured interview), we show \textit{how grandparents and grandchildren perform intergenerational storytelling without technology support in terms of content, presentation, and organization}; and 2) RQ2: Based on the part 2 of our study (i.e., AR probe study and co-design workshop), we show \textit{ways in which AR could facilitate photo-based intergenerational storytelling in these three aspects}.} We annotated the grandparent, grandchild, and participant group as \textbf{GP}, \textbf{GC}, and \textbf{G}, respectively.




\subsection{\revision{How Grandparents and Grandchildren Perform Photo-based Intergenerational Storytelling}}
\revision{We present our findings about the ways how grandparents and grandchildren perform photo-based intergenerational storytelling in their daily lives focusing on the three key aspects of storytelling: Content, Presentation, and Organization.}
\subsubsection{Content}
We identified three types of frequently mentioned content in the intergenerational stories and the characteristics that participants expressed for each type of content.

\textbf{Content 1: Family portraits and photos of family members' activities}. Family portraits and photos of family members' activities are the most prominent type of content mentioned by participants in their intergenerational stories. Old photos selected by GPs were about family portraits (N=4) or their activities (N=6) by one or more family members. For stories about family portraits, GPs usually introduced the characters' clothes, hairstyles, and their relationships with the GC since the GC went away to study, moved, and went abroad, gradually losing the impression of this person. GPs wanted their GCs to \textit{``get to know these people again''(GP9)} through intergenerational storytelling. The activities in the photos included a trip, an outing, a family dinner, etc. GPs usually described the location of the activity, the time, the name, and the functions of the building involved (e.g., train station, airport, museum, etc.), and some obsolete daily items (e.g., an old suitcase, a Kerosene lamp, film rolls, etc.).


\textbf{Content 2: Changes in technology, society, people, and artifacts}. ‘Changes' was another frequent point of interest for GPs and GCs. `Changes' in technology, society, people, and artifacts were mentioned almost in every intergenerational dyad (N=9). These `changes' sit at the nexus of the younger and the older generation, which brought creative discussion and motivation for storytelling. For example, GPs \revision{(N=7)} often mentioned changes in photography technology and societal changes with the aim of seeking identification that their grandchildren could understand despite these faded, black-and-white photos were not very exquisite and beautiful for the young, they still have their precious value for GPs. GP7 introduced photo-taking experiences when they were young, which were taken in a photo studio when they did not own a camera and afterward got a physical photo which was expensive: \textit{`` taking photos in the past tended to commemorate something after a well-thought-out decision, unlike now just pressing the shutter at will.''}. Furthermore, GCs \revision{(N=6)} expressed interest and curiosity in \textit{``Changes of their parents' life experiences''}. A relatively long discussion would often be triggered by this topic where the GP introduced more details about the growth of the GC's parents (i.e., their sons or daughters) that the GC may not know earlier. During this process, GPs tended to talk about the upbringing experiences \revision{(GP1)}, clothes, and appearances of the GC's parents \revision{(GP2, GP10)}. This sharing was often followed by GCs' sharing of their impressions of their parents \revision{(N=3)}. Such sharing about GCs' parents from both GPs and GCs often brought conflicting views about GCs' parents. While GCs \revision{ (GC1, GC2, GC5)} tended to use words like \textit{"strict," "executive," and "gentle,"}, GPs \revision{ (GP1, GP2, GP5)} tended to use \textit{"naughty" or "upwardly mobile"} instead. Such conflicting views of the same person in the family are often a common point of interest in the storytelling process for both GPs and GCs.

\textbf{Content 3: Descriptions to enhance the accuracy of story elements.}All GPs expressed a strong desire to give an accurate description of things that they mentioned during storytelling. Many GPs (N=8) kept revising and qualifying descriptions and tried to recall the details of the old time as much as possible, with the hope that \revision{\textit{``my grandchildren could understand the story elements, such as an old item'' (GP6)}}, that GCs did not see or experience, especially when they encountered some commemorative for a family or representative cultural items. The reason might be GPs want to express the proper knowledge or culture for their GCs as illustrated by GP10's quotes when he was asked for the reasons for explaining ``Nian Hua'' (Chinese ancient Spring Festival paintings):
\begin{quote}
    \textit{``I always try to polish and qualify my languages when communicating with my grandchild. Because I want them to be able to have a realistic imagination of the past time. Furthermore, the obsolete items I mentioned that the stories often still have some cultural attributes, so it is important to describe them correctly and responsibly to help younger people not forget them.'' }
\end{quote}

\subsubsection{Presentation}
We identified two tendencies in how GPs presented their story narrations.

\textbf{Presentation 1: GPs tended to present their stories verbally and rarely incorporated \revision{facial expressions, hand gestures, or variations to their tones}}. 
\revision{The most common mode of storytelling observed was oral narrations by a monotonous and straightforward delivery, without more vivid presentation elements as illustrated by GCs such as the lacking of \textit{``facial expressions'' (GC2)}, \textit{``gestures'' (GC8, GC9)}, \textit{``variations to tones'' (GC6, GC10)}. Consequently, some of the GCs reported \textit{``easily feeling bored'' (GC2, GC10)}, \textit{``easily getting distracted'' (GC6, GC8)}, or \textit{``being unwilling to make proactive interactions'' (GC8)} with GPs during their storytelling.}
\revision{Moreover, instead of incorporating prompting questions, GPs primarily relied on statements in their stories. Specifically, in 7 participant groups, questions were mainly proposed by GCs first and engaged in no more than five interactions overall, with each interaction consisting of no more than three dialogue rounds.} \revision{Consequently, \revision{GCs expressed the expectation for GPs to use richer story presentation approaches, such as \textit{``using more descriptive gestures''(GC6)}, \textit{``adding prompting questions''(GC9)} and \textit{``adding variations in voice intonation''(GC6)}}, to catch their interest}, as illustrated by the following quote:
\begin{quote}
    ``I hope my grandpa can narrate the story in a more vivid manner, similar to my teacher in class, by actively posing questions and using gestures to illustrate the mentioned items.'' - GC9
\end{quote}



\textbf{Presentation 2: GPs tended to have difficulty finding words that were both accurate enough to express their thoughts and familiar enough to GCs for them to understand.} Many GPs (N=7) found it hard to find the right words to describe an item or an event that their GCs did not see or experience. Despite this difficulty, all GPs made efforts to illustrate the concept of obsolete items in or related to the target photo by offering more verbal descriptions. GCs' questions also prompted GPs to seek different ways to describe their ideas, which sometimes could confuse GCs even more: many GCs (N=6) asked clarification questions about the described item but remained confused even after hearing more verbal descriptions. One main reason was the mismatch of knowledge between GPs and GCs. For example, GP5 mentioned \textit{``Gong Fen''} (credits of one's social workload) and \textit{``Hong Hui Zheng''} (a special type of credential), which were the proper terms in China's Proletarian Cultural Revolution. It was out of GC5's knowledge. Therefore, GC5 was confused and repeatedly asked for further illustration of its appearance and usage. However, GP5 stated, \revision{\textit{``I found it hard to explain such concept and its usage in this short period of time, as these things are not widely adopted in the present age and even probably being phased out. ''}} This unresolved information hindered the story, sapping the enthusiasm of both sides of the storytelling.

\subsubsection{Organization} 
\label{sec:organization}
\begin{figure}[bth!]
    \centering
    \includegraphics[width=0.8\textwidth]{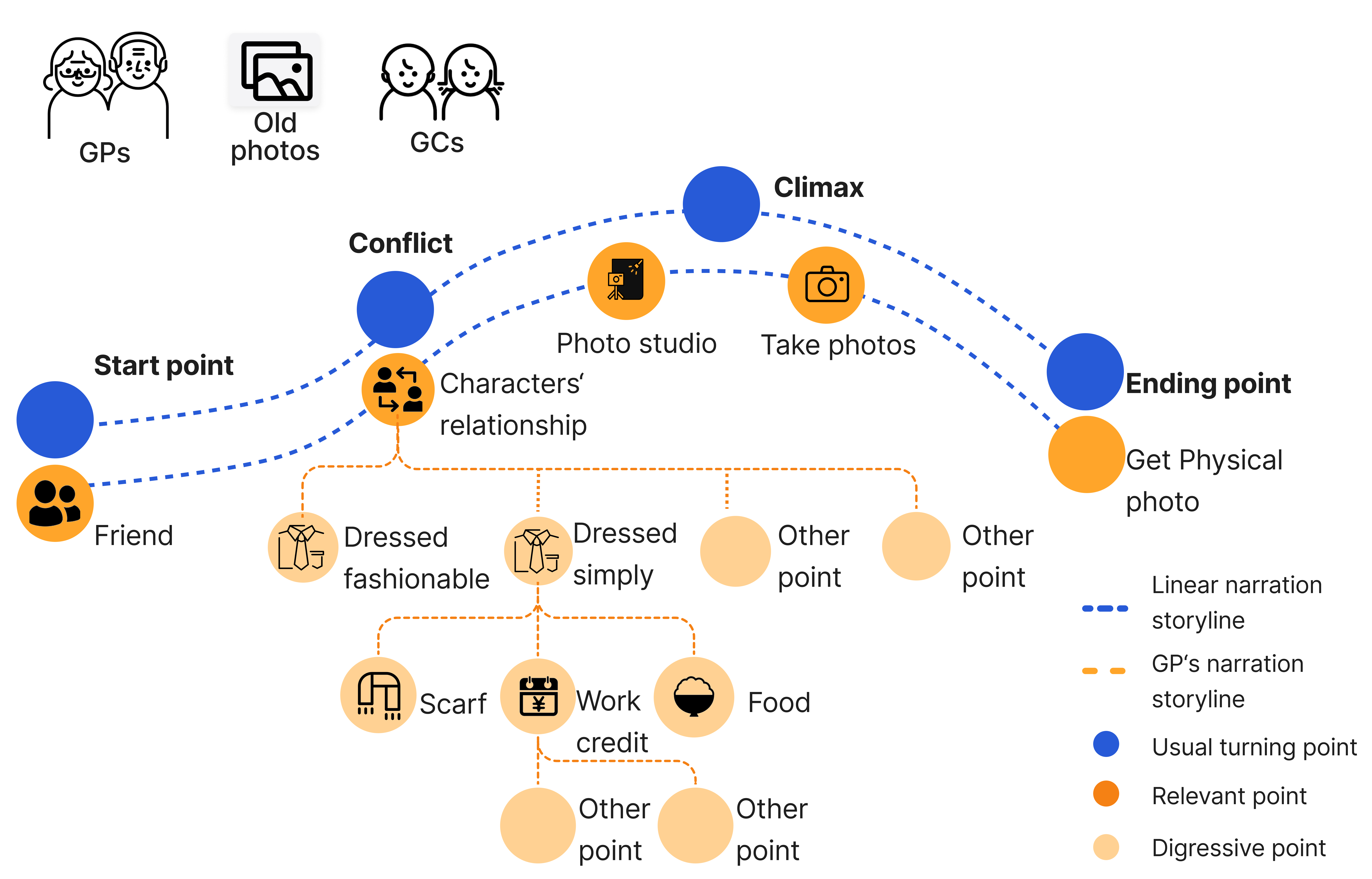}
    \caption{An example storyline from GP5's story. The blue dotted lines refer to the main storyline of the story, which is a linear structure constituted with different narrative stages (blue nodes), connecting all the relevant events. The orange dotted lines refer to GP's narration storyline and the nodes connected by them formed a tree structure with digressive points.}
    \Description{This figure illustrated our findings in how GPs organized their stories.}
    \label{fig:organization}
\end{figure}

Organization refers to the ways how GPs organize their stories. We present the organization approaches of GPs and the perception of GCs for these organization approaches. Fig. \ref{fig:organization} shows the connections of the findings on this topic.

\textbf{Organization 1: GPs did not always stay on the main storyline and may diverge from it to add supplemental information, which may further perpetuate them away from the main storyline.} A large number of GPs (N=7) narrated the story in a divergent tree structure rather than a linear structure that continuously developed the main storyline. As illustrated in Fig. \ref{fig:organization}, typically, they started from an element in the photo and iteratively introduced supplementary information (e.g., the usage of an item) and side stories. As a result, such a narrative structure generated many branches and child nodes of side stories that gradually detached from the main storyline. For example, as shown in Fig. \ref{fig:organization}, GP5 started by telling a story about taking photos with his friend (on the blue dotted main thread). When he talked about the characters' relationships, he gradually drifted away from the main thread and talked about the \textit{``scarf''}, and then \textit{''work credits''}, which are digressive nodes on orange dotted lines:
\begin{quote}
    \textit{"I went to a photo studio to take photos with a good friend of mine. I dressed simply. He was from the city, and his family was richer than mine. He dressed very fashionably. I wore a scarf on my head. Hey, do you know what a scarf is? It is different from the scarf you know. It was used by farmers to wipe the sweat on their heads. At that time, farmers worked hard every day to get some work credits. You must not know what work credits were. For example, it was a kind of credit for people's production activities at that time. Only when you earned work credits through labor, you could use them to exchange for food.  At that time, there was no concept of money, and we ..." -GP5} 
\end{quote}
\revision{Consequently, GC5 still did not learn about the purpose of the photo and what happened after the photo was taken while getting knowledge of the \textit{``scarf''}, and then \textit{''work credits''}. GC5 expressed, \textit{``While getting to know these less relevant items is interesting, I still want my grandpa to focus more on the main idea of the story.''}}
While side stories from the main storyline might be interesting and trigger interactions between GPs and GCs, most GCs (N=6) still expressed the hope for a more compact and clear structure in GPs' stories to get a better understanding of the main storyline.



\textbf{Organization 2: GPs preferred to link their stories about an old photo to a related person than other elements (e.g., time, location).}
GPs tended to remember an old photo by linking it to a related person. Instead of using the time or location related to an old photo, many GPs (N=6) preferred to use a related person to organize and tell their stories about the photo. Furthermore, one common way to organize stories was first developing the story from the related person and describing the person's related information (e.g., the age and the relationship). As shown in Fig. \ref{fig:organization}, the start point of the story is the \textit{``Friend''} node in the GP5's narration storyline. A common pattern used by GPs (N=9) was as follows: \textit{``This is the photo of [a person]...''}. Additionally, GPs commonly mentioned the relationships of the people in old photos, including the generational relationship of the people and their relationships with GPs. \revision{GPs found the relationships were important because \textit{``The people in the photo were my childhood buddies, and they hold great significance in my life. However, my grandchild might not be familiar with them. I really wish that through sharing my story, my grandchild can get to know them better.''(GP7)}}


\subsection{Ways in Which AR Could Facilitate Photo-based Intergenerational Storytelling for GPs and GCs}

From the participatory design workshops, we identified seven ways in which AR could help to augment photo-based intergenerational stories in terms of the content, presentation, and organization reported in the previous section. While the five ways are about assisting GPs to better tell their stories, the last two are about better engaging GCs and GPs into the stories.




\begin{figure}[tbh!]
    \centering
    \includegraphics[width=0.8\textwidth]{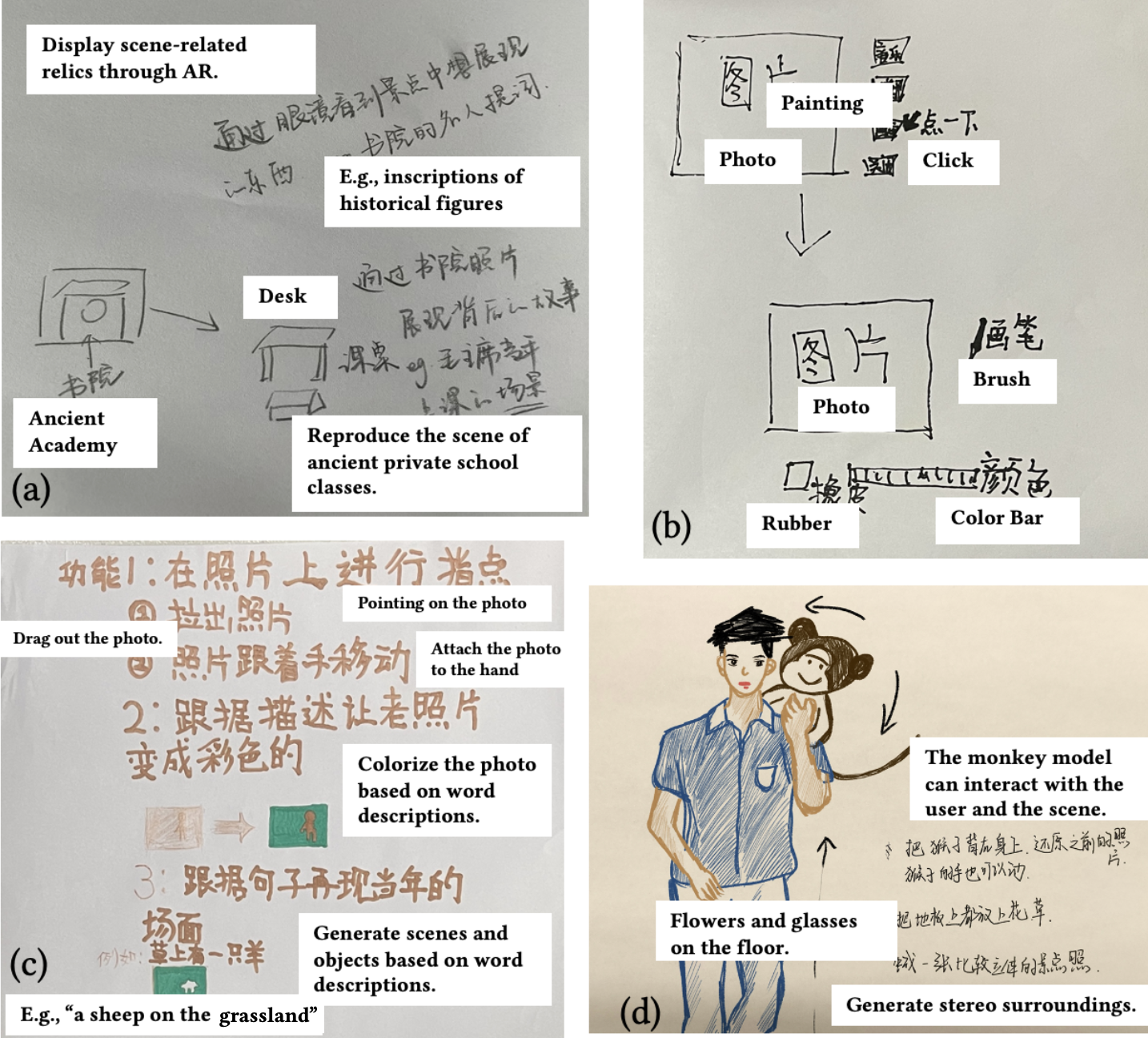}
    \caption{Examples of users' sketches from the participatory design. Key descriptions and paragraphs were annotated in English.}
    \Description{This figure demonstrates four examples of users' sketches from the participatory design. Key descriptions and paragraphs were annotated in English.}
    \label{fig:design_examples}
\end{figure}

\textbf{AR could augment story elements with digital content (e.g., 3D models) related to the old photo to help GPs better explain relevant artifacts, events, or persons and for GCs to visually comprehend them.} All the participants favored the idea of showing 3D models of an element in the photo by \textit{Probe 3 (Display of 3D models)} in Fig. \ref{fig:probes}, which could help communicate difficult-to-articulate information between GPs and GCs to save GPs' efforts in revising their descriptions (Content 3) or finding the proper vocabulary to express their ideas (Presentation 2) as illustrated by the following quote:
\begin{quote}
     ``3D models in AR could help me easily show the details in the old photo, it saves me a lot of time to clarify features of items I want to show. And when I talked with my grandchild, I always want to present as much detail as possible to make the story more interesting.'' - GP6
 \end{quote}
 Moreover, participants also proposed more complex elements, such as scenes related to the old photo. As shown in Fig. \ref{fig:design_examples} (a), GC10 mentioned the possibility of rendering the scene of ancient private classes with inscriptions of historical figures, desks, and the ancient academy. Participants further proposed more scenes to be displayed by AR 3D models, such as some special venues in the photo, the square of the train station, the lake in the park, etc., making people feel that \textit{``we can 'step into' the story in the photo.'' (GP9)}


\textbf{AR could support GPs and GCs to edit the augmented content (e.g., 3D model) to better match their story elements.} With current AI technologies, it is possible to perform speech recognition on GPs' stories and use the keywords mentioned to search AR assets libraries and find relevant content (e.g., 3D models) to augment their stories. However, the augmented content may not match the ones in GPs' minds. Since GPs tended to spend efforts revising their story elements to deliver accurate descriptions (Content 3), they expressed the need to be able to edit the augmented content and offered two possible approaches to do so. The first one was to restore older photos. 
All groups approved the idea of the \textit{Probe 6 (Photo Restoration) in Fig. \ref{fig:probes}} and would like to restore black-and-white into well-colorized ones or restore worn photos into complete ones. Meanwhile, instead of relying on automatic restoration algorithms (e.g., colorization) that are still limited in their ability to fully restore photos, some GCs proposed to let their GPs make further modifications to the initially restored photos by algorithms so that they could better tune the modifications to meet their expectations. GC1 proposed the concrete modification tool, as is shown in Fig. \ref{fig:design_examples} (b), which is an AR palette tool including the brush, the rubber, and the color bar. Moreover, GPs could modify restored photos by adjusting the color (G7, G8, G10), texture (G3, G5), and filters (G2, G5, G10), etc., to restore the damage that occurred over time. The second approach was to provide appropriate options for modifications instead of asking GPs to sketch freely in the air as current AR interactions often do. Many GPs (N=6) mentioned they were not good at drawing with their bare hands on paper, let alone in the middle air. Instead, they preferred to have a set of possible options to choose from and perhaps accept suggested modifications, such as changing colors and fine-tuning shapes.


\textbf{AR could allow GPs to leverage physical objects and AR-augmented similar items to explain relevant story elements.}
After trying \textit{Probe 5 (Reminiscence)} in Fig. \ref{fig:probes}, GPs could see the physical old photo in their hands being displayed in AR and also generated a similar photo which is also the beach, including a mother and a daughter, as is shown in Fig \ref{fig:probes}. All the GPs and GCs were supportive of leveraging physical old photos to help generate AR-augmented similar items mentioned in the story to help GPs conveniently compare the old photo with similar new photos to clarify the changes in technology, society, people, and artifacts (Content 2). Moreover, GPs proposed to leverage more daily physical objects (e.g., cups, paper, clothes, etc.) to help explain similar items in old times in AR. It could provide references or alternative editable models in AR for GPs to describe items. For example, GP8 mentioned an old times cup (\textit{``Tangci Bei''}) in the story, he described that \textit{"I could grasp the cup on the tea table, then the AR system could generate the \textit{``Tangcibei Bei''} model automatically, then I could use some modification tools in AR to edit the model such as modifying the shape, the size, etc. to explain it to my grandchild.''} It reduced the difficulty of GPs to finding proper words to explain the items to GCs (Presentation 2).

\textbf{AR could enable flexible perspective shifts to reduce the burden of manipulating the augmented content.} When telling a story about a character or an item (Content 1), GPs often need to manipulate the augmented AR content (e.g., 3D models) to a suitable location for their GCs to observe. Frequent dragging and rotating the AR content \revision{at a distance} caused an extra burden for GPs. For example, as shown in Fig. \ref{fig:design_examples} (c), GP1 commented, \textit{``I wish all the objects could be attached to my hand and follow my hand to move so that I do not need to spend a long time dragging them every time.''}. Although our probes provided the ability for GPs to manipulate AR content, only one GP (GP4) learned to use \textit{Probe 1 (Pointing)} in Fig. \ref{fig:probes} quickly, and other GPs \revision{experienced failures when using the virtual ray and cursor to point objects at a distance.} Although all the GCs grasped this probe more quickly, they still \revision{needed} some time to get accustomed to using the virtual cursor to point. Therefore, participants proposed to introduce more flexible and automated perspective shifts in AR that can save them a lot more effort and trigger more proactive interaction (Presentation 1) between GPs and GCs since the GC can be more easily aligned with the GP's perspective, which \textit{``leads to fewer distractions and more immersive feelings'' (GC8).}

\textbf{AR could provide indicators (e.g., annotation timeline, keywords) for GPs to keep track of their main storyline and side stories so that they could tell stories in a more organized way.} 
As explained in Sec. ~\ref{sec:organization}, it was common that GPs to diverge from the main storyline to talk about side stories and had difficulty resuming the main storyline (Organization 1). Consequently, listeners (GCs) might get lost when following the story. To address the issue, participants proposed that AR could provide some visual indicators (e.g., keywords shown on a sticky note, annotations shown on a timeline) in the field of views of GPs so that they could better keep track of the main storyline and resume easily when needed.  
For example, AR could show the keywords that GPs have already talked about and perhaps the frequency of these keywords being mentioned in a timeline. Such visual indicators could not only help GPs keep track of their story but also help GCs understand the story in particular when many new terms were mentioned that they might not necessarily remember all. 
For example, GC2 mentioned that \textit{"Grandma talked about the characteristics of the clothes of the past era, but I couldn't remember them well. If some annotations could appear in AR as a prompt like those used in a classroom learning session, it would make me more impressed."}

\textbf{AR could enable augmented content to be interactive to better engage GCs in the storytelling process.} As explained earlier, GCs might feel bored when hearing their GPs tell their stories without active participation. Thus, a large number of GCs (N=8) expressed a strong desire to interact with the AR-augmented objects or characters mentioned in their GPs' stories. As shown in Figure \ref{fig:design_examples} (d), GP3 and GC3 suggested that models be more interactive, \textit{``for example, the monkey model could climb on people, because at that time, monkeys did climb up my shoulder, and I was so scared!''}. GP4 mentioned  \textit{``I hope I could hand over the clothes model to my grandchild in AR so that he could try it on and better appreciate the style of clothing at the time. Moreover, I can take a picture of myself wearing the clothes as a souvenir and make more proactive interactions with my grandchild!''} GCs (N=4) also mentioned that AR could offer more collaboration opportunities for both them and their GPs to collectively interact with the 3D models. Furthermore, making the story characters (Organization 2) interactive could potentially enable proactive interactions (Presentation 2) between GPs and GCs. For example, GPs could teach GCs how to greet the character in the story in traditional ways or let the character perform some actions to visually show their personality.

\textbf{AR could leverage multi-sensory channels, including vision, hearing, and olfactory, to provide more engaging experiences of the story for GPs and GCs.} Most participant groups (N=8) felt that the sound provided by \textit{Probe 4 (Sound effects)} in Fig. \ref{fig:probes} helped GCs better appreciate the visual content of the photo, such as a lively wet market. Inspired by this sound effect probe, GPs and GCs further proposed multi-sensory channels, including haptic and olfactory to their storytelling. For olfactory, GP7 mentioned that \textit{"I wish AR could provide the smell of the wet market, which was really difficult for me to describe (Presentation 2) so that my grandchild could really feel the wet market and be more interested in my story."} For the hearing channel, participants further proposed to use spatial audio, dialect dialogue assets, and voice interaction. As is shown in Fig. \ref{fig:design_examples} (d), GC3 proposed to generate stereo surroundings, such as importing spatial audio to better illustrate the environment of the park. GP5 mentioned that some dialect dialogues and their translation into the standard spoken language could also be considered since many older adults were more comfortable or even only speaking dialects. Such features would potentially help GPs express themselves comfortably and also help their GCs, who may not fully understand local dialects appreciate their GPs' stories better. Moreover, using dialect dialogues that matched the characters' backgrounds in the storytelling process could potentially make these characters more vivid (Organization 2), as one GP explained, \textit{"My grandchildren could have a better understanding of the character I mentioned." } Many GPs also hoped to use their local dialects to evoke 3D content in AR. Furthermore, voice interaction was mentioned by most participants (N=7). As is shown in Fig. \ref{fig:design_examples} (c), GC1 and GP1 proposed that their verbal descriptions could be used to generate visual scenes or to restore old photos (e.g., colorization).

\section{Discussion}
We first present the key takeaways of our research and our key contributions in Section 6.1, and then we further discuss the design considerations based on our findings in Section 6.2. 

\subsection{Key Takeaways}
We identified how grandparents and grandchildren perform photo-based
intergenerational storytelling in terms of story content, presentation, and organization. Our findings concluded three types of commonly-mentioned content: 1) family portraits and photos of family members' activities, 2) changes in technology, society, people, and artifacts, and 3) descriptions to enhance the accuracy of story elements. We identified two tendencies of presenting content: 1) GPs tended to present their stories verbally and rarely used visual or
audio support, and 2) GPs tended to have difficulty in finding words that were accurate enough to express their thoughts and familiar enough for GCs to understand; Finally, we identified two common ways of organizing stories: 1) GPs did not always stay on the main storyline and may diverge from it
to add supplemental information, and 2) GPs preferred to link their stories of an old photo to a related person than other elements (e.g., time, location).

We also uncovered several ways in which AR could be leveraged to facilitate the storytelling process. 
At a high level, our findings show that GPs had a strong motivation to tell stories with old photos to their GCs and wanted to make their story content accurate, engaging, and understandable. For example, when telling stories, GPs regularly revised their descriptions of story elements and had difficulty finding accurate words that were also understandable to GCs. When exploring the role of AR in design workshops, GPs felt that AR-augmented content (e.g., 3D models) could help them enrich their story content and alleviate these issues. Nevertheless, they still expressed the desire to be able to edit AR content to better match their anticipation. This suggests that AR can offer augmented content to help with storytelling and points out the need to design intuitive tools for GPs to perform effective AR content editing.

In terms of the organization of their stories, GPs tended to deviate from their main storyline to tell side stories and had difficulty telling their stories in an organized way that did not cause confusion for GCs. While previous work pointed out that it is hard for improvised oral storytelling in a family to be organized \cite{papacharissi2016affective, liem2020structure}, our research focused on photo-based intergenerational storytelling between GPs and GCs and offered insights into how they might leverage AR to address this challenge, such as using augmented visual indicators (e.g., timeline annotation, keywords) to keep track of the storyline. 

Furthermore, our work highlights the importance of interaction and shows several ways in which AR could help enrich the interaction between GPs and GCs. For example, our findings show that collaborative interactions should be augmented by AR in enriching the storytelling process. While previous work identified collaborative interactions could be conducive to enhancing the relationship between the two sides \cite{10.1145/2212776.2223698,doi:10.1080/03637750500322453,doi:10.1080/15267430903401441}, our work uncovered several ways in which GPs and GCs could interact with the story in AR, such as empowering the interactability of AR models, enabling perspective shifts, and introducing multi-sensory channels to improve the proactive interactions between GPs and GCs.



\subsection{Design Implications and Considerations}

\subsubsection{Content Generation and Editing}

Our findings revealed that GPs were sensitive to the preciseness of their story elements (e.g., characters, items, and scenes). Although AR is capable of visualizing content with better vividness, expressiveness, and immersion, it remains a challenge how best to design tools to allow GPs to create and edit content effectively in AR. 
One approach is to have an AR asset database that contains 3D models of common characters, items, architectures, and pictures that might likely appear in GPs' stories. These assets can be indexed by descriptive tags (e.g., in terms of category, shape, color, usage, religion, and age) so that they can be retrieved by a querying system, for example, using natural language (e.g., large language models such as GPT-3\cite{brown2020language}). 
Moreover, our findings suggested that GPs and GCs wanted to have more personalized content, this approach should also allow them to add their personalized content, such as family photos and scanned item copies, for preservation and sharing (e.g., used in storytelling) purposes. 

Another approach is to generate content with generative AI models (e.g., Stable Diffusion \cite{rombach21diffusion} and Magic3D \cite{lin22magic3d}) that take users' verbal input as text prompts and synthesize images and 3D models at high resolution. However, as our findings showed that GPs still wanted to be able to edit the generated content as it might not match what they had in mind, future work should investigate how to enable GPs, perhaps GCs as well, to effectively and iteratively edit the generated content.  

Toward this goal, human-AI collaborated approaches can be explored. For example, an AI-assisted editing system guided by text (speech) with auxiliary input channels (e.g., strokes and real-world referents) could be implemented based on large language models and diffusion generative models \cite{kawar22imagic,nichol21glide,yang22paint}. In terms of interaction experience, GPs could simply say \textit{``change the photo to a 60-century Chinese style''}, or \textit{``change the roof of this house into a red one'' (while circling the house region with the finger)} to make corresponding modifications. GPs can also edit by referring to a real-world object, \textit{``The kettle is like that one'' (pointing at a real-world kettle), ``but more worn-out and with a bigger handle''}. Moreover, prompt-based editing could be designed to allow GPs to edit the content iteratively until their expectations are met. 


\subsubsection{Organization and Flexibility of Intergenerational Storytelling}
As reported in Sec. \ref{sec:organization}, our findings revealed that GPs needed some help to better organize their storytelling process. 
Through our design workshop, GPs and GCs proposed that some form of indicators (e.g., timelines annotation) in AR could potentially help them keep track of their stories.  
Such visual indicators could also be instrumental for GCs to better understand the story elements. This finding is in line with prior work that suggested that audio, pictures, and symbols can act as links to build shared understanding in storytelling \cite{10.1145/3555158,10.1145/2971485.2996732}. However, it remains an open question of how best to implement hints in the context of photo-based intergenerational storytelling between GPs and GCs. Specifically, what annotation should be presented to GPs? When and how should they be presented to help GPs and GCs keep track of the story without distracting them from it? Are annotations the same for GPs and GCs? 

Although GPs and GCs expressed a desire for an organized structure in storytelling, intergenerational storytelling is often improvised, and some form of \textit{flexibility} in the story structure can be beneficial too \cite{VASALOU2020100214,10.1007/978-3-319-48279-8_7}. For example, improvised oral storytelling relaxes the requirement that actions are strictly logical and allow the narrator to not follow a linear story structure \cite{10.1007/978-3-319-48279-8_7}. Furthermore,  previous research suggested that side stories (stories that are not related to the main topic) are not all unacceptable to listeners in improvised oral storytelling, especially in a family setting \cite{doi:10.1080/15267430903401441}. Our findings also found that side stories could trigger more topics between intergenerational dyads and make the story more interesting. Therefore, one interesting direction is to investigate how to balance between maintaining the organization of the main story and introducing side stories to keep GPs and GCs engaged. 

\subsubsection{Enable Collaborative Interactions between GPs and GCs}
Storytelling between parents and children is a collaboration process requiring the participation of both storytellers and story listeners \cite{10.1145/3415169}, and collaborative work between them can be conducive to their relationship \cite{10.1145/3479585}. Extending this literature to storytelling between GPs and GCs, our findings found that there were opportunities in the intergenerational storytelling process for GPs and GCs to collaborate. For example, GPs and GCs could collaboratively manipulate and edit augmented content (e.g., 3D models), which might trigger interactions and conversations between them. 

Furthermore, another way to enable collaboration between GPs and GCs in the intergenerational storytelling process is to gamify it. Previous work suggested that collaborative games could facilitate intergenerational communication \cite{10.1145/3415169,siyahhan2018families} between parents and children \cite{golsteijn2013facilitating}, grandparents and grandchildren \cite{10.1007/s00779-020-01364-9}. Specifically, collaborative games could promote symmetrical interactions and create a comfortable and enjoyable atmosphere for intergenerational storytelling \cite{10.1145/3415169}. The study of intergenerational learning (IGL) also stressed that social interactions should not be a one-way process but should benefit both generations \cite{ropes2013intergenerational}. Future work should investigate ways to enable a more reciprocal relationship between GPs and GCs. For example, while GPs could take the role of conveying cultural values and wisdom from their life experiences,  GCs could serve to provide explanations and guide older adults to better experience possibilities enabled by AR technologies.

\subsection{Limitations and Future Work}
Our work investigated \revision{how grandparents and grandchildren perform photo-based intergenerational storytelling} and took a first step to explore how AR might be able to assist this process through a mixed-method study of storytelling, interview, and AR probes-enabled co-design workshop. While our findings offer insights into their storytelling practices and seven ways how AR could be designed, our work has limitations that motivate potential future research directions.

First, we conducted the study with a limited number of GP-GC pairs recruited from urban communities in China, where our research team all resided. The backgrounds of our participants (e.g., intimacy, age groups, gender, prior storytelling experience, and culture) might have some effects on our findings. For example, as the stories shared by GPs were related to photos and their life experiences, such content could be different from country to country. While we believe that our findings revealed common needs and challenges of intergenerational storytelling and the role of AR in it, more studies with grandparents and grandchildren from other countries are needed to investigate the effect of cultures and extend our findings. 
In addition, as the age of GCs ranged from 12 to 22, there might be differences for them to engage in intergenerational storytelling. From our observation, older GCs (>18 years old) were more considerate and patient, actively providing guidance on using AR headsets for GPs than younger GCs. However, due to our limited samples, we do not highlight this finding. Future work with larger samples should be conducted to investigate how the age of GPs might affect the storytelling process. 


Second, although our AR probe-based co-design workshop revealed interesting insights, it lasted for a relatively short period. One interesting open question is how GPs and GCs might use AR for storytelling in longer terms. It would be interesting to first implement AR-based intergenerational storytelling based on our findings and design implications and then investigate how GPs and GCs would actually use them in their storytelling process for a longer time, for example, through a deployment study. 


Lastly, we noted that all GPs in our study played the role of the storyteller while GCs of story listeners. Future work could investigate the practices and challenges of other forms of storytelling (e.g., GCs as the storyteller) or other forms of interaction between them (e.g., storytelling with games, quizzes, and competitions). For example, the GP and the GC take turns telling their stories and making quizzes for each other. \revision{ Furthermore, we chose to use one AR device in our study (one AR device for the storyteller and one PC screen for the listener) to reduce the complexity of the setup and focus on exploring the possibility of AR in helping their storytelling process, but we agree that using two AR devices is also a possible approach to augmenting storytelling (e.g., how to deal with conflicts on the same target in AR during the storytelling process, how to manage the authority of both parties to operate the targets during storytelling, etc.). Hence, the roles of the storyteller and listener may not remain fixed, allowing for fluid exchanges of storytelling between grandparents and grandchildren. }

\section{Conclusion}
We have investigated photo-based intergenerational storytelling between grandparents and grandchildren in terms of content, presentation, and organization through an in-person storytelling observation and a semi-structured interview and explored ways in which AR might assist their intergenerational storytelling processes in these three aspects through a participatory design workshop with AR probes. We designed 6 AR probes to solicit design ideas from intergenerational pairs for leveraging AR to enrich their storytelling processes. Our findings revealed the possible ways of intergenerational storytelling. Specifically, we concluded three types of commonly-mentioned content, two tendencies of presenting content, and two common ways of organizing stories. Our work also revealed seven ways in which AR can enhance it. In sum, our work took a first step to identify the opportunities in photo-based intergenerational storytelling between grandparents and grandchildren and highlighted opportunities and design implications for AR designers and researchers who are interested in promoting social interaction between grandparents and grandchildren through AR-based storytelling.

\begin{acks}
This work is partially supported by the Guangzhou Science and Technology Program City-University Joint Funding Project (Project No. 2023A03J0001) and Guangdong Provincial Key Lab of Integrated Communication, Sensing and Computation for Ubiquitous Internet of Things.
\end{acks}

\bibliographystyle{ACM-Reference-Format}
\bibliography{main}

\appendix

\end{document}